\documentclass[a4paper, 12pt]{article}
\usepackage{amsmath, amssymb}
\usepackage{graphicx}
\usepackage{subfigure}
\usepackage{verbatim}
\usepackage{cite}
\usepackage[left=20mm, top=20mm, right=20mm, bottom=25mm]{geometry}

\renewcommand{\vec}[1]{{\boldsymbol{#1}}}
\DeclareMathOperator{\re}{Re}
\DeclareMathOperator{\im}{Im}

\DeclareMathOperator{\tr}{Tr}

\begin{document}
\title{Probability representation of kinetic equation for open quantum system}
\author{V. I. Man'ko$^1$, V. A. Sharapov$^2$, E. V. Shchukin$^1$\\
\small $^1$P.N. Lebedev Physical Institute, Leninskii Prospekt 53, Moscow 119991, Russia\\
\small $^2$Institute of Spectroscopy, Troitsk, 142092 Moscow reg., Russia}
\date{}
\maketitle

\begin{abstract}
The tomographic probability distribution is used to decribe the kinetic equations for open
quantum systems. Damped oscillator is studied. Purity parameter evolution for different
damping regime is considered.
\end{abstract}

\section{Introduction}

There exist several formulations of quantum mechanics (see recent review in
\cite{ajp-70-288}). In all these formulations the state of a system is associated with
mathematical structure (complex wave function, density operator, path integral) which is
different from standard probability distribution used in classical statistics. On the other
hand there exists a possibility of specific representation of quantum mechanics used in recent
works \cite{jrlr-22-401, pl-a-309-176} where the quantum state is described by the standard
probability distribution (tomogram or tomographic symbol of density operator). This
representation of quantum states is called "probability representation". This representation
can be used both for conservative and open quantum systems. In this work we focuse on the open
quantum systems. We review the approach and clarify some technical methods to deal with the
tomograms of observables (operators). The tomograms are generalized functions. In view of this
the calculations with tomograms need some specific attention.

We will study quantum kinetic equations which describe damping in quantum domain
(\cite{np-a-645-376, jmp-34-3887, tp-167-7, tp-191-171}). Evolution of conservative quantum
system is described by Schr\"{o}dinger equation for wave function \cite{ap-79-489}. The
quantum systems interacting with a heat bath (open systems) obey to quantum kinetic equations
written for density operator introduced in \cite{zp-45-430,
neumann-mathematische-grundlagen-quantenmechanik}. The density operator is the hermitian
operator with nonnegative eigenvalues. The diagonal matrix elements of the density operator in
arbitrary basis have the meaning of standard probability distribution. In view of this the
trace of density operator has to be equal unity. Quantum kinetic equations must to be close in
such form that the described general properties of the density operator have to be preserved
in the process of quantum evolution. It means that the initial density operator
$\hat{\rho}(0)$ is transformed into the density operator $\hat{\rho}(t)$ using a specific map.
This map $\hat{\rho}(0) \to \hat{\rho}(t)$ has to preserve nonnegativity, hermiticity and
trace of the density operator. Such kind of map was introduced by Sudarshan \cite{pr-121-920}
for finite-dimensional Hilbert space and it was studied in \cite{pr-121-920, cjm-24-520,
kraus-states-effects-operations}. The most general quantum kinetic equations which satisfy the
demands to preserve the properties of density operator in the process of time evolution have
been suggested in \cite{rpm-3-247, cmp-46-43, cmp-48-119}. The evolution equation for damped
harmonic oscillator was used in \cite{jqe-3-348}. Recently \cite{pl-a-213-1, fp-27-801} new
probability representation of quantum mechanics was introduced. In this representation the
quantum state is described by standard positive probability distribution (tomographic
probability distribution). In the probability representation the Schr\"{o}dinger (and von
Newmann) evolution equation takes the classical-like form of equation for the standard
probability distribution of Fokker-Planck type. On the other hand till now the general quantum
kinetic equations for open systems have not been studied in the probability representation.
The aim of our work is to obtain the kinetic equations preserving the properties of density
operator in the process of time evolution in the probability representation of quantum
mechanics. In fact we will use the specific procedure of tomographic star-product quantization
\cite{jp-a-35-699} and geometric approach to tomographic map \cite{jp-a-34-8321} which
generalize the Moyal \cite{pcps-45-99} star-product quantization. In the Moyal approach the
quantum kinetic equation is written for Wigner quasidistribution function \cite{pr-40-749}. In
the tomographic approach the kinetic equation is written for the standard positive probability
distribution function. It is worthy to note that some nonlinear dynamical equations were
considered recently \cite{jo-b-5-95} in the tomographic representation. The equations of
quantum mechanics are linear equations though there are some attempts to introduce nonlinear
equations describing quantum state evolution (see e.g. \cite{ap-237-226}). The linear generic
quantum kinetic equations for open systems can be also presented in the form of Fokker Planck
type equation as well.

The paper is organized as follows. In next Section 2 we discuss some properties of tomograms
as generalized functions. In Section 3 we review tomographic representation of quantum
mechanics. In Section 4 we derive the tomographic form of quantum evolution equation. In
Section 5 example of damped oscillator is given. Conclusion are presented in section 6.

\section{Tomographic symbol of the unity operator}

Let us consider the tomogram of unity operator. We check the validity of the obvious equality
\begin{equation}
    w_{\hat{1}} \ast w_{\hat{1}} = w_{\hat{1}}.
\end{equation}
This equality follows from operator equality
\begin{equation}
    \hat{1} \cdot \hat{1} = \hat{1}.
\end{equation}
It can be shown that the tomographic symbol of the unity operator reads
\begin{equation}
    w_{\hat{1}}(X,\mu,\nu) = -\pi|X|\delta(\mu)\delta(\nu).
\end{equation}
The tomogram $w_{\hat{A}}(X, \mu, \nu)$ is a probability distribution if $\hat{A}$ is a
density operator. For arbitrary operator $\hat{A}$ its tomogram can be negative and even
complex function. We have to check the equality
\begin{equation}
\begin{split}
   &\frac{1}{{(2\pi)}^2} \int \pi^2|X_1||X_2|
    \delta(\mu_1)\delta(\nu_1)\delta(\mu_2)\delta(\nu_2)
    \delta\Bigl(\mu(\nu_1+\nu_2)-\nu(\mu_1+\mu_2)\Bigr) \times \\
   &\exp\left\{\frac{i}{2}\left[\nu_1\mu_2-\nu_2\mu_1+2(X_1+X_2)-
    \left(\frac{\nu_1+\nu_2}{\nu}+
    \frac{\mu_1+\mu_2}{\mu}\right)X\right]\right\}\,
   dX_1\,d\mu_1\,d\nu_1\,dX_2\,d\mu_2\,d\nu_2 \\
   &= -\pi|X|\delta(\mu)\delta(\nu).
\end{split}
\end{equation}
First of all, we take the factors containing the variables $X_1$ and $X_2$:
\begin{equation}
\begin{split}
   &\frac{1}{4} \int \left( \int |X|e^{iX}\,dX \right)^2
    \int \delta(\mu_1)\delta(\nu_1)\delta(\mu_2)\delta(\nu_2)
    \delta\Bigl(\mu(\nu_1+\nu_2)-\nu(\mu_1+\mu_2)\Bigr) \times \\
   &\exp\left\{\frac{i}{2}\left[\nu_1\mu_2-\nu_2\mu_1-
    \left(\frac{\nu_1+\nu_2}{\nu}+\frac{\mu_1+\mu_2}{\mu}\right)X
    \right]\right\}\,d\mu_1\,d\mu_2\,d\nu_1\,d\nu_2.
\end{split}
\end{equation}
Taking into account the equality \cite{gelfand-shilov}
\begin{equation}
    \int |X|e^{iX}\,dX = -2,
\end{equation}
we can reduce the equality under study to the following one
\begin{equation}\label{e-ReducedTomUnit}
\begin{split}
   &\int \delta(\mu_1)\delta(\nu_1)\delta(\mu_2)\delta(\nu_2)
    \delta\Bigl(\mu(\nu_1+\nu_2)-\nu(\mu_1+\mu_2)\Bigr) \times \\
   &\exp\left\{\frac{i}{2}\left[\nu_1\mu_2-\nu_2\mu_1-
    \left(\frac{\nu_1+\nu_2}{\nu}+\frac{\mu_1+\mu_2}{\mu}\right)X
    \right]\right\}\,d\mu_1\,d\mu_2\,d\nu_1\,d\nu_2 \\
   &= -\pi|X|\delta(\mu)\delta(\nu).
\end{split}
\end{equation}
To evaluate this integral we make the change of variables defined by the formulas
\begin{equation}
\begin{aligned}
    \mu_1 + \mu_2 &= l\mu, \qquad & \nu_1 + \nu_2 &= k\nu, \\
    \mu_1 - \mu_2 &= \chi, \qquad & \nu_1 - \nu_2 &= \lambda.
\end{aligned}
\end{equation}
The inverse relations read
\begin{equation}
\begin{aligned}
    \mu_1 &= \frac{1}{2}(l\mu + \chi), \qquad & \nu_1 &= \frac{1}{2}(k\nu + \lambda), \\
    \mu_2 &= \frac{1}{2}(l\mu - \chi), \qquad & \nu_2 &= \frac{1}{2}(k\nu - \lambda).
\end{aligned}
\end{equation}
The Jacobian related to the introduced linear transformation has the form
\begin{equation}
    \frac{\partial(\mu_1,\mu_2,\nu_1,\nu_2)}{\partial(l,\chi,k,\lambda)} = \frac{1}{4}|\mu\nu|.
\end{equation}
Taking into account the equality
\begin{equation}
    \delta\Bigl(\mu(\nu_1+\nu_2)-\nu(\mu_1+\mu_2)\Bigr) =
    \frac{1}{|\mu\nu|}\delta\left(\frac{\nu_1+\nu_2}{\nu}-\frac{\mu_1+\mu_2}{\mu}\right)
\end{equation}
we can transform the left-hand side of the equality \eqref{e-ReducedTomUnit} to the following
form
\begin{equation}
\begin{split}
   &\int\delta(l\mu+\chi)\delta(l\mu-\chi)\delta(k\nu+\lambda)
    \delta(k\nu-\lambda)\delta(k-l)\times\\
   &\times\exp\left\{\frac{i}{2}\left[\frac{1}{2}(l\lambda\mu-k\chi\nu)-(k+l)X
    \right]\right\}\,dk\,dl\,d\lambda\,d\chi = \\
   &\int\delta(l\mu+\chi)\delta(l\mu-\chi)\delta(k\nu+\lambda)\delta(k\nu-\lambda)
    \exp\left\{\frac{i}{4}k(\lambda\mu-\chi\nu)-ikX\right\}\,dk\,d\lambda\,d\chi = \\
   &\int\delta(2k\mu)\delta(2k\nu)\exp\left\{\frac{i}{4}k(k\nu\mu-k\mu\nu)-ikX\right\}\,dk =
    -\frac{\pi}{4}|X|\delta(\mu)\delta(\nu).
\end{split}
\end{equation}
It means that we get the desired result.

We also check the validity of the equality
\begin{equation}
    w_{\psi} \ast w_{\psi} = w_{\psi}.
\end{equation}
for the tomogram $w_\psi$ of a pure state. This equality follows from the property
\begin{equation}
    \hat{\rho}_{\psi}^2 = \hat{\rho}_{\psi}
\end{equation}
of density operator $\hat{\rho}_{\psi}$ of normalized pure state. The tomographic symbol
$w_{\psi}(X,\mu,\nu)$ of the density operator $\hat{\rho}_{\psi}$ is described by the formula
(see e.g. \cite{jp-a-35-699})
\begin{equation}
\begin{split}
    w_{\psi}(X,\mu,\nu) &= \frac{1}{2\pi|\nu|}{\int\left|\psi(y)
    \exp\left\{\frac{i\mu}{2\nu}y^2-i\frac{X}{\nu}y\right\}\,dy\right|}^2 = \\
   &\frac{1}{2\pi|\nu|}\int\psi(y)\psi^*(z)
    \exp\left\{\frac{i\mu}{2\nu}(y^2-z^2)-i\frac{X}{\nu}(y-z)\right\}\,dy\,dz.
\end{split}
\end{equation}
According to the formulas for star-product kernel given in \cite{jp-a-35-699} the star-product
$w_{\psi} \ast w_{\psi}$ takes the form:
\begin{equation}\label{e-ww}
\begin{split}
   &(w_{\psi} \ast w_{\psi})(X,\mu,\nu) =
    \frac{1}{16\pi^4}\int\frac{1}{|\nu_1\nu_2|}
    \psi(y_1)\psi^*(z_1)\psi(y_2)\psi^*(z_2) \times \\
   &\exp\left\{\frac{i\mu_1}{2\nu_1}(y_1^2-z_1^2)+\frac{i\mu_2}{2\nu_2}(y_2^2-z_2^2)-
    i\frac{X_1}{\nu_1}(y_1-z_1)-i\frac{X_2}{\nu_2}(y_2-z_2)\right\}
    \delta\Bigl(\mu(\nu_1+\nu_2)-\nu(\mu_1+\mu_2)\Bigr) \times \\
   &\exp\left\{\frac{i}{2}\left[\nu_1\mu_2-\nu_2\mu_1+2(X_1+X_2)-
    \left(\frac{\nu_1+\nu_2}{\nu}+
    \frac{\mu_1+\mu_2}{\mu}\right)X\right]\right\}\,
    dy_1\,dz_1\,dy_2\,dz_2\,dX_1\,d\mu_1\,d\nu_1\,dX_2\,d\mu_2\,d\nu_2.
\end{split}
\end{equation}
Choosing the factors containing the variables $X_1$ and $X_2$ and using the equality
\begin{equation}
    \int\exp\left\{i\left(1-\frac{y_1-z_1}{\nu_1}\right)X_1\right\}
    \exp\left\{i\left(1-\frac{y_2-z_2}{\nu_1}\right)X_2\right\}\,dX_1\,dX_2 =
    4\pi^2|\nu_1\nu_2|\delta(\nu_1-y_1+z_1)\delta(\nu_2-y_2+z_2)
\end{equation}
we reduce the expression \eqref{e-ww} to the following one
\begin{equation}
\begin{split}
   &(w_{\psi} \ast w_{\psi})(X,\mu,\nu) =
    \frac{1}{4\pi^2} \int\psi(y_1)\psi^*(z_1)\psi(y_2)\psi^*(z_2) \times \\
   &\exp\left\{\frac{i\mu_1}{2\nu_1}(y_1^2-z_1^2)+\frac{i\mu_2}{2\nu_2}(y_2^2-z_2^2)\right\}
    \delta(\nu_1-y_1+z_1)\delta(\nu_2-y_2+z_2)
    \delta\Bigl(\mu(\nu_1+\nu_2)-\nu(\mu_1+\mu_2)\Bigr) \times \\
   &\exp\left\{\frac{i}{2}\left[\nu_1\mu_2-\nu_2\mu_1-
    \left(\frac{\nu_1+\nu_2}{\nu}+\frac{\mu_1+\mu_2}{\mu}\right)X
    \right]\right\}\,dy_1\,dz_1\,dy_2\,dz_2\,d\mu_1\,d\nu_1\,d\mu_2\,d\nu_2.
\end{split}
\end{equation}
Integrating the variables $\nu_1$ and $\nu_2$ we get the expression
\begin{equation}
\begin{split}
   &(w_{\psi} \ast w_{\psi})(X,\mu,\nu) =
    \frac{1}{4\pi^2} \int\psi(y_1)\psi^*(z_1)\psi(y_2)\psi^*(z_2)
    \exp\left\{\frac{i}{2}\mu_1(y_1+z_1)+\frac{i}{2}\mu_2(y_2+z_2)\right\} \times \\
   &\delta\Bigl(\mu(y_1-z_1+y_2-z_2)-\nu(\mu_1+\mu_2)\Bigr)
    \exp\left\{\frac{i}{2}\left[(y_1-z_1)\mu_2-(y_2-z_2)\mu_1- \right.\right. \\
   &\left.\left.\left(\frac{y_1-z_1+y_2-z_2}{\nu}+\frac{\mu_1+\mu_2}{\mu}\right)X
    \right]\right\}\,dy_1\,dz_1\,dy_2\,dz_2\,d\mu_1\,d\mu_2.
\end{split}
\end{equation}
To obtain the result we make the change of variables defined by the formulas
\begin{alignat}{2}
    \mu_1 + \mu_2 &= k, \qquad & \mu_1 &= \frac{1}{2}(k+l), \\
    \mu_1 - \mu_2 &= l, \qquad & \mu_2 &= \frac{1}{2}(k-l).
\end{alignat}
The Jacobian related to the change of variables reads
\begin{equation}
    \frac{\partial(\mu_1,\mu_2)}{\partial(k,l)} = \frac{1}{2}.
\end{equation}
One can transform the expression for the star-product $w_{\psi} \ast w_{\psi}$ to the
following form
\begin{equation}
\begin{split}
   &(w_{\psi} \ast w_{\psi})(X,\mu,\nu) =
    \frac{1}{8\pi^2} \int\psi(y_1)\psi^*(z_1)\psi(y_2)\psi^*(z_2)\times\\
   &\times\exp\left\{\frac{i}{4}(k+l)(y_1+z_1)+\frac{i}{4}(k-l)(y_2+z_2)\right\} \times \\
   &\delta\Bigl(\mu(y_1-z_1+y_2-z_2)-k\nu\Bigr)
    \exp\left\{\frac{i}{2}\left[\frac{1}{2}(y_1-z_1)(k-l)-
    \frac{1}{2}(y_2-z_2)(k+l)-\right.\right. \\
   &\left.\left.\left(\frac{y_1-z_1+y_2-z_2}{\nu}+\frac{k}{\mu}\right)X\right]\right\}\,
    dy_1\,dz_1\,dy_2\,dz_2\,dk\,dl.
\end{split}
\end{equation}
Taking into account the equality
\begin{equation}
    \delta\Bigl(\mu(y_1-z_1+y_2-z_2)-k\nu\Bigr) =
    \frac{1}{|\nu|}\delta\Bigl(k-\frac{\mu}{\nu}(y_1-z_1+y_2-z_2)\Bigr)
\end{equation}
and integrating over the variable $k$ we get
\begin{equation}
\begin{split}
    (w_{\psi} \ast w_{\psi})(X,\mu,\nu) &=
    \frac{1}{8\pi^2|\nu|} \int\psi(y_1)\psi^*(z_1)\psi(y_2)\psi^*(z_2)
    \exp\left\{\frac{i\mu}{2\nu}(y_1-z_1+y_2-z_2)(y_1+z_2)+ \right. \\
   &\left.\frac{i}{2}l(z_1-y_2)-i\frac{X}{\nu}(y_1-z_1+y_2-z_2)\right\}\,
    dy_1\,dz_1\,dy_2\,dz_2\,dl.
\end{split}
\end{equation}
Using the Fourier representation of delta-function
\begin{equation}
    \int\exp\left\{\frac{i}{2}l(z_1-y_2)\right\}\,dl = 4\pi\delta(z_1-y_2),
\end{equation}
we obtain the equality
\begin{equation}\label{e-eq}
\begin{split}
    (w_{\psi} \ast w_{\psi})(X,\mu,\nu) &=
    \frac{1}{2\pi|\nu|} \int\psi(y_1)\psi^*(z_1)\psi(y_2)\psi^*(z_2)
    \exp\left\{\frac{i\mu}{2\nu}(y_1-z_1+y_2-z_2)(y_1+z_2) \right. \\
   &\left.-i\frac{X}{\nu}(y_1-z_1+y_2-z_2)\right\}\delta(z_1-y_2)\,dy_1\,dz_1\,dy_2\,dz_2 = \\
   &\frac{1}{2\pi|\nu|} \int\psi(y_1)\psi^*(y_2)\psi(y_2)\psi^*(z_2)
    \exp\left\{\frac{i\mu}{2\nu}(y_1^2-z_2^2)-i\frac{X}{\nu}(y_1-z_2)\right\}\,
    dy_1\,dy_2\,dz_2.
\end{split}
\end{equation}
Taking into account the normalization of the wave function $\psi(x)$
\begin{equation}
    \int{|\psi(x)|}^2\,dx = 1,
\end{equation}
we get the desired equality
\begin{equation}
\begin{split}
    (w_{\psi} \ast w_{\psi})(X,\mu,\nu) = \frac{1}{2\pi|\nu|} \int\psi(y)\psi^*(z)
    \exp\left\{\frac{i\mu}{2\nu}(y^2-z^2)-i\frac{X}{\nu}(y-z)\right\}\,dy\,dz =
    w_{\psi}(X,\mu,\nu).
\end{split}
\end{equation}
If we take the arbitrary density operator $\hat{\rho}$ instead of the density operator
$\hat{\rho}_{\psi}$ of pure state, we see that all formulas will be correct except the last
one. Making corresponding substitution $\psi(y_1)\psi^*(z_1)\psi(y_2)\psi^*(z_2) \to
\rho(y_1,z_1)\rho(y_2,z_2)$ we get
\begin{equation}
    (w_{\hat{\rho}} \ast w_{\hat{\rho}})(X,\mu,\nu) =
    \frac{1}{2\pi|\nu|} \int\rho(y_1,y_2)\rho(y_2,z_2)
    \exp\left\{\frac{i\mu}{2\nu}(y_1^2-z_2^2)-i\frac{X}{\nu}(y_1-z_2)\right\}\,
    dy_1\,dy_2\,dz_2.
\end{equation}
One can see that the following relation take place
\begin{equation}
    w_{\hat{\rho}} \ast w_{\hat{\rho}} = w_{\hat{\rho}}\quad \Leftrightarrow\quad
    \rho(x,x^{\prime}) = \int\rho(x,y)\rho(y,x^{\prime})\,dy\quad \Leftrightarrow\quad
    \hat{\rho}^2=\hat{\rho}.
\end{equation}
Thus, we proved the identity for the tomograms of pure quantum states.

\section{Tomographic representation of quantum mechanics}

In this section we review the tomographic approach given in \cite{pl-a-213-1, fp-27-801}. In
\cite{qso-7-615} an operator $\hat{\vec{X}} = (\hat{X}_{1},....,\hat{X}_{N})$ is discussed for
the case of N = 1 as a generic linear combination of the position and momentum operators
$\hat{X}_{n} = \mu_{n}\hat{x}_{n} + \nu_{n}\hat{p}_{n}$, where $\mu_{n}$ and $\nu_{n}$ are
real parameters for $n=1,...,N$, and $\hat{\vec{X}}$ is Hermitian, hence observable. The
physical meaning of $\vec{\mu} = (\mu_{1},...,\mu_{N})$ and $\vec{\nu} =
(\nu_{1},...,\nu_{N})$ is that they describe an ensemble of rotated and scaled reference
frames, in classical phase space, in which the position $\vec{X}$ may be measured. It was
shown \cite{qso-7-615} that the quantum state of a system is completely determined if the
classical probability distribution $w(\vec{X},\vec{\mu},\vec{\nu})$, for the variable
$\vec{X}$ is given in an ensemble of reference frames in the classical phase space. Such a
function, also known as the marginal distribution function or quantum tomogram, belongs to a
broad class of distributions which are determined as the Fourier transform of a characteristic
function \cite{pr-177-1882}.  The following formula for a quantum tomogram was derived in
\cite{jp-a-34-8321}:
\begin{equation}\label{E:two}
    w(\vec{X},\vec{\mu},\vec{\nu}) =
    \langle\delta(\vec{X}-\vec{\mu}\hat{\vec{q}}-\vec{\nu}\hat{\vec{p}})\rangle,
\end{equation}
where $\langle\hat{A}\rangle=\tr(\hat{\rho}\hat{A})$ is the average value of the operator
$\hat{A}$ on the state described by the density operator $\hat{\rho}$. In \cite{pr-177-1882}
it was shown that, whenever $\hat{\vec{X}}$ is observable, $w(\vec{X},\vec{\mu},\vec{\nu})$ is
indeed a probability distribution, as it is positive definite and satisfies the normalization
condition $\int w(\vec{X},\vec{\mu},\vec{\nu})\,d\vec{X} = 1$. It is worthy to note that in
view of homogenity of Dirac delta-function the tomogram has the same homogenity property.
Connection between the density matrix $\rho(\vec{x},\vec{x}')$ and quantum tomogram
$w(\vec{X},\vec{\mu},\vec{\nu})$ may be expressed through the following relation
\cite{pl-a-213-1}:
\begin{equation}\label{E:six}
    \rho(\vec{x},\vec{x}') = \frac{1}{(2\pi)^{N}}\int w(\vec{Y},\vec{\mu},\vec{x}-\vec{x}')
    \prod_{n=1}^{N}\exp\left[i\left(Y_{n}-\mu_{n}\frac{x_{n}+x'_{n}}{2} \right) \right]\,
    d\vec{\mu}\,d\vec{Y}.
\end{equation}

\section{Evolution equation in tomographic representation}

In this Section we introduce the quantum kinetic equations in tomographic form. The general
evolution equation for the density matrix of the open system was suggested in \cite{rpm-3-247,
cmp-48-119, cmp-46-43, rmp-13-149}. Solution of this equation corresponds to the Hermitian
nonnegative operator with constant trace at all moments of time. According to
\cite{cmp-48-119} the most general evolution equation which keeps stated properties takes the
form:
\begin{equation}
\label{e:Lind} \dot{\widehat{\rho}} = -i\left[\widehat{H},\widehat{\rho}\right] +
                 \frac{1}{2}\sum\limits^{n}_{j=1}\left(2\widehat{V}_j\widehat{\rho}\widehat{V}^+_j-
                 \widehat{V}^+_j\widehat{V}_j\widehat{\rho}-\widehat{\rho}\widehat{V}^+_j\widehat{V}_j\right),
\end{equation}
where $\hat{H}$ is Hamiltonian, $\hat{V}_j$ are arbitrary linear operators and $n$ is an
arbitrary integer. This equation, presented in coordinate representation, has the form:
\begin{equation}\label{e:2}
\begin{split}
    \dot{\rho}\left(\vec{x},\vec{x}^{\prime},t\right) =
    &\Biggl[-i\Bigl(H\left(\vec{x},\vec{p}\right)-
    H\left(\vec{x}^{\prime},-\vec{p}^{\prime}\right)\Bigr)+
    \frac{1}{2}\sum\limits^n_{j=1}\Bigl(2V_j\left(\vec{x},\vec{p}\right)
    V_j\left(\vec{x}^{\prime},-\vec{p}^{\prime}\right)-\Bigr.\Biggr. \\
    &-\Biggl.\Bigl.V_j\left(\vec{x},-\vec{p}\right)
    V_j\left(\vec{x},\vec{p}\right)-
    V_j\left(\vec{x}^{\prime},-\vec{p}^{\prime}\right)
    V_j\left(\vec{x}^{\prime},-\vec{p}^{\prime}\right)\Bigr)\Biggr]
    \rho\left(\vec{x},\vec{x}^{\prime},t\right),
\end{split}
\end{equation}
where $\vec{p}=-i\partial/\partial{\vec{x}}$,
$\vec{p}^{\prime}=-i\partial/\partial{\vec{x}^{\prime}}$. The kinetic equation in such a form
can be easily transformed into equation on quantum tomogram.  Indeed, the density matrix in
coordinate representation can be expressed through tomogram $w(\vec{X},\vec{\mu},\vec{\nu})$
using relation \eqref{E:six}. Therefore we can establish the following correspondence of the
action of the operators on the density matrix $\rho\left(\vec{x},\vec{x}^{\prime}\right)$ and
the marginal distribution $w(\vec{X},\vec{\mu},\vec{\nu})$ \cite{jrlr-18-407}:
\begin{equation}
\label{e:p}
\begin{split}
    \vec{x}                                     &\mapsto
        \vec{Q}      = -{\left(\frac{\partial}{\partial\vec{X}}\right)}^{-1}\frac{\partial}{\partial\vec{\mu}} +
                       \frac{i}{2}\vec{\nu}\frac{\partial}{\partial{\vec{X}}}\qquad
    \vec{x}^{\prime}                            \mapsto
        \vec{Q}^{\prime} = -{\left(\frac{\partial}{\partial\vec{X}}\right)}^{-1}\frac{\partial}{\partial\vec{\mu}} -
                       \frac{i}{2}\vec{\nu}\frac{\partial}{\partial\vec{X}}\\
  \vec{p}          &\mapsto
        \vec{P}          = -\frac{i}{2}\vec{\mu}\frac{\partial}{\partial\vec{X}} -
                           {\left(\frac{\partial}{\partial\vec{X}}\right)}^{-1}\frac{\partial}{\partial\vec{\nu}}\qquad
    \vec{p}^{\prime}  \mapsto
        \vec{P}^{\prime} = -\frac{i}{2}\vec{\mu}\frac{\partial}{\partial\vec{X}} +
                           {\left(\frac{\partial}{\partial\vec{X}}\right)}^{-1}\frac{\partial}{\partial\vec{\nu}}
\end{split}
\end{equation}
One can see that all the operators in right hand side of above formulas have scale invariant,
i.e. they are not changed if we replace $X_i \to \lambda_iX_i$, $\mu_i \to \lambda_i\mu_i$,
$\nu_i \to \lambda_i\nu_i$. Thus using \eqref{e:2} and \eqref{e:p} one gets what we call
evolution equation in tomographic (or probability) representation
\begin{equation}\label{e-w}
\begin{split}
    \dot{w}(\vec{X},\vec{\mu},\vec{\nu},t) &=
    \Biggl[-i\Bigl(H\left(\vec{Q},\vec{P}\right)-H\left(\vec{Q^{\prime}},-\vec{P}^{\prime}\right)\Bigr)+
    \frac{1}{2}\sum\limits^n_{j=1}\Bigl(
    2V_j\left(\vec{Q},\vec{P}\right)V_j\left(\vec{Q}^{\prime},-\vec{P}^{\prime}\right)-\Bigr.\Biggr.\\
    &-\Biggl.\Bigl.V_j\left(\vec{Q},-\vec{P}\right)
    V_j\left(\vec{Q},\vec{P}\right)-V_j\left(\vec{Q}^{\prime},-\vec{P}^{\prime}\right)
    V_j\left(\vec{Q}^{\prime},-\vec{P}^{\prime}\right)\Bigr)\Biggr]
    w(\vec{X},\vec{\mu},\vec{\nu},t),
\end{split}
\end{equation}
where operators $\hat{\vec{Q}}$, $\hat{\vec{Q}}^{\prime}$, $\hat{\vec{P}}$ and
$\hat{\vec{P}}^{\prime}$ act on $w(\vec{X},\vec{\mu},\vec{\nu})$ according to \eqref{e:p}. Due
to scale invariance of \eqref{e:p} the evolution equation preserves the homogenity property of
tomogram.

\section{Partial case of one-dimensional evolution equation in probability representation}\label{s:2}

There exist several examples of kinetic equation for open systems. In this section we consider
the tomographic evolution equation for the case of one-dimensional oscillator with Hamiltonian
$\hat{H}$ of the form
\begin{equation}\label{e:ham1}
    \hat{H}=\frac{\hat{p}^2}{2}+\frac{\hat{x}^2}{2}
\end{equation}
(we assume that $\hbar = 1$, $m=1$, $\omega = 1$) and the only operator $V(\hat{q},\hat{p}) =
u\hat{x}+v\hat{p}$, taken as a linear combination of the operators $\hat{x}$ and $\hat{p}$
with arbitrary complex coefficients $u$ and $v$. Some aspects of tomographic representation
for this oscillator were considered in \cite{pl-a-213-1}. For $u=\sqrt{\gamma/2}$ and
$v=i\sqrt{\gamma/2}$ we have the standard quantum kinetic equation
\begin{equation}
\dot{\widehat{\rho}} = -i\left[\widehat{H},\widehat{\rho}\right] +
                 \frac{\gamma}{2}\sum\limits^{n}_{j=1}\left(2\widehat{a}_j\widehat{\rho}\widehat{a}^+_j-
                 \widehat{a}^+_j\widehat{a}_j\widehat{\rho}-\widehat{\rho}\widehat{a}^+_j\widehat{a}_j\right).
\end{equation}
For such a choice of operators $\hat{H}$ and $V(\hat{q},\hat{p})$ equation \eqref{e-w} for the
tomogram $w(X,\mu,\nu,t)$ can be represented in the following way:
\begin{equation}\label{e:part}
\begin{split}
    \dot{w}(X,\mu,\nu,t)&=\left[\mu\frac{\partial}{\partial{\nu}}-
    \nu\frac{\partial}{\partial{\mu}}+
    \frac{1}{2}(|u|^2\nu^2 + |v|^2\mu^2)\left(\frac{\partial}{\partial{X}}\right)^2+ \right. \\
   & \left.+ \im{(uv^*)}\left(\mu\frac{\partial}{\partial{\mu}}+
    \nu\frac{\partial}{\partial{\nu}}\right) -
    \re{(uv^*)}\mu\nu\left(\frac{\partial}{\partial{X}}\right)^2 \right]w(X,\mu,\nu,t).
\end{split}
\end{equation}
To solve this equation we make the Fourier transformation
\begin{equation}
    \tilde{w}(k,\mu,\nu)=\frac{1}{2\pi}\int w(X,\mu,\nu)\exp{(-ikX)}\, dX
\end{equation}
and obtain the following equation for Fourier components $\tilde{w}(k,\mu,\nu,t)$ of the
tomogram $w(X,\mu,\nu,t)$:
\begin{equation}\label{e:Shrod}
    \dot{\tilde{w}}(k,\mu,\nu,t)=-i\left(\frac{1}{2}\hat{\vec{\xi}}^T\Gamma\hat{\vec{\xi}}-
    i\im{(uv^*)}\right)\tilde{w}(k,\mu,\nu,t),
\end{equation}
where $4$-vector $\hat{\vec{\xi}}=(\hat{p}_{\mu},\hat{p}_{\nu},\hat{x}_{\mu},\hat{x}_{\nu}) =
(-i\partial/\partial\mu,-i\partial/\partial\nu,\mu,\nu)$ and $4 \times 4$ symmetric matrix
$\Gamma$ is of the form
\begin{equation}\label{e:bb1}
    \Gamma=\begin{Vmatrix}
          \Gamma_{pp}& \Gamma_{px} \\
          \Gamma_{xp}& \Gamma_{xx}
      \end{Vmatrix}=
      \begin{Vmatrix}
          0 & 0 & -\im(uv^*) & 1 \\
          0 & 0 & -1 & -\im(uv^*) \\
          -\im(uv^*) & -1 & -i|v|^2k^2 & i\re(uv^*)k^2 \\
          1 & -\im(uv^*)  & i\re(uv^*)k^2 & -i|v|^2k^2
      \end{Vmatrix}.
\end{equation}
Equation \eqref{e:Shrod} for the Fourier components $\tilde{w}(k,\mu,\nu,t)$ is Shr\"{o}dinger
type equation with effective Hamiltonian given by a quadratic form. Equations of such a type
can be solved in the framework of time dependent invariants method
\cite{malkin-manko-dynamical-symmetries-coherent-states-quantum-systems}. According to this
method, Green function $G(\mu,\nu,\mu',\nu',t)$ of equation \eqref{e:Shrod}, i.e. function
which connects $\tilde{w}(k,\mu ,\nu , t)$ and $\tilde{w}(k,\mu,\nu, 0)$ by the equality
\begin{equation}
\tilde{w}(k,\mu,\nu,t)=\int G(\mu,\nu,\mu',\nu',t) \tilde{w}(k,\mu',\nu',0)\,d\mu'\,d\nu',
\end{equation}
can be obtained by solving the system of linear differential equations on $2\times 2$ matrices
$\Lambda_{1}$, $\Lambda_{2}$, $\Lambda_{3}$, $\Lambda_{4}$:
\begin{equation}\label{e:de1}
\begin{aligned}
   \dot{\Lambda}_1 &= \Lambda_1 \Gamma_{xp} - \Lambda_2 \Gamma_{pp},
   \qquad &
   \dot{\Lambda}_3 &= \Lambda_3 \Gamma_{xp} - \Lambda_4 \Gamma_{pp},       \\
   \dot{\Lambda}_2 &= \Lambda_1 \Gamma_{xx} - \Lambda_2 \Gamma_{px},
   \qquad &
   \dot{\Lambda}_4 &= \Lambda_3 \Gamma_{xx} - \Lambda_4 \Gamma_{px},
\end{aligned}
\end{equation}
with initial conditions $\Lambda_{1}=E_2$, $\Lambda_{2}=0$, $\Lambda_{3}=0$,
$\Lambda_{4}=E_2$, where $E_2$ is $2\times 2$ unity matrix. The solution of this system for
the matrix $\Gamma$ given by \eqref{e:bb1} reads
\begin{equation}\label{e:lam}
\begin{aligned}
    \Lambda_{1} &=
    \begin{Vmatrix}
        \cos{t} & -\sin{t}\\
        \sin{t} & \cos{t}
    \end{Vmatrix}
    e^{-2\im(uv^*)t}, & \qquad
    \Lambda_{3} &=
    \begin{Vmatrix}
        0 & 0 \\
        0 & 0
    \end{Vmatrix}, \\
    \Lambda_{4} &=
    \begin{Vmatrix}
        \cos{t} & -\sin{t}\\
        \sin{t} & \cos{t}
    \end{Vmatrix}
    e^{2\im(uv^*)t}, & \qquad
    \Lambda_{2} &=
    2
    \begin{Vmatrix}
        a  & b \\
        b & -c
    \end{Vmatrix}
    \sinh\Bigl(-\im(uv^*)t\Bigr)\cos{t} + \\
   & & &+\begin{Vmatrix}
        -2b & a+c \\
        a+c & 2b
    \end{Vmatrix}
    \cosh\Bigl(-\im(uv^*)t\Bigr)\sin{t} + \\
   & & &+\begin{Vmatrix}
        0 & c-a \\
        a-c & 0
    \end{Vmatrix}
    \sinh\Bigl(-\im(uv^*)t\Bigr)\sin{t},
\end{aligned}
\end{equation}
where constants $a$, $b$ and $c$ are determined by
\begin{equation}
\begin{split}
    a &= i\frac{2|v|^2\im^2(uv^*)+2\re(uv^*)\im(uv^*)+|u|^2+|v|^2 }{4\im(uv^*)\Bigl(1+\im^2(uv^*)\Bigr)}k^2, \\
    b &= -i\frac{2\re(uv^*)\im(uv^*)+|u|^2-|v|^2 }{4(1+\im^2(uv^*))}k^2, \\
    c &= -i\frac{|u|^2\im^2(uv^*)-4\re(uv^*)\im(uv^*)+2|u|^2+2|v|^2 }{8\im(uv^*)\Bigl(1+\im^2(uv^*)\Bigr)}k^2.
\end{split}
\end{equation}
For the case $\im(uv^*)=0$ the constant $a$ and $c$ are not defined but in the expression for
matrix $\Lambda_2$ remains the only term in which these constants appear in the combinations
$a+c$ and $a-c$. It is easy to see that these combinations are defined even if $\im(uv^*)=0$.
The Green function $\tilde{G}(\mu,\nu,\mu',\nu',t)$ of equation \eqref{e:Shrod} for the case
of $\Lambda$-matrices \eqref{e:lam} is determined by formula
\cite{malkin-manko-dynamical-symmetries-coherent-states-quantum-systems}:
\begin{equation}
    \tilde{G}(\mu,\nu,\mu',\nu',t)=\frac{1}{\sqrt{\det\Lambda_{4}}}
    \delta\Bigl((\mu,\nu)-\Lambda_{1}(\mu',\nu')\Bigr)\exp{\left(-\frac{i}{2}(\mu',\nu')\Lambda_{2}\Lambda_{1}\begin{pmatrix}\mu' \\ \nu'\end{pmatrix}\right)}.
\end{equation}
Let us consider the evolution of the coherent state of the system with Hamiltonian
\eqref{e:ham1}. Notice, that obtained Green function allows consideration of evolution of an
arbitrary state. The tomogram corresponding to the coherent state of the oscillator is (see
e.g. \cite{pr-a-57-3291})
\begin{equation}\label{e:o}
    w_{\alpha} (X,\mu,\nu,0)=\frac{1}{\sqrt{\pi (\mu^2 +\nu^2)}}
    \exp{\left(-\frac{\Bigl[X- \sqrt{2}\re{(\alpha)}\mu-
    \sqrt{2}\im{(\alpha)}\nu\Bigr]^2}{\mu^2 +\nu^2} \right)}.
\end{equation}
Taking the Fourier components of the tomogram $w_{\alpha}(X,\mu,\nu,t)$ one can find the
evolution of $\tilde{w}_{\alpha}(k,\mu,\nu,t)$. After performing inverse Fourier transform of
$\tilde{w}_{\alpha}(k,\mu,\nu,t)$ we obtain the evolution of the tomogram
$w_{\alpha}(X,\mu,\nu,t)$ of the form:
\begin{equation}\label{e:wco}
    w_{\alpha}(X,\mu,\nu,t)=\frac{1}{\sqrt{\pi (C\mu^2 + D\nu^2 + E\mu\nu)}}
    \exp{\left(-\frac{[X-\lambda\mu-\delta\nu]^2}
    {C\mu^2 + D\nu^2 + E\mu\nu } \right)},
\end{equation}
where $\lambda=\langle\hat{x}\rangle$, $\delta=\langle\hat{p}\rangle$, $C=\sigma_{xx}$,
$D=\sigma_{pp}$ and $E=2\sigma_{xp}$. Considering evolution of the coherent state, one should
distinguish two cases:  $\im(uv^*) \neq 0$ and $\im(uv^*)=0$. The coefficients $C=C(u,v,t)$,
$D=D(u,v,t)$ and $E=E(u,v,t)$ reads
\begin{equation}\label{e:c1c2}
\begin{aligned}
      & \text{For the case of}\ \im(uv^*) \neq 0 & & \text{For the case of}\ \im(uv^*) = 0 \\
    C &= \Bigl(1 - d + c\cos{2t} - e\sin{2t}\Bigr)\times  & \qquad
         C &= 1+(|u|^2 + |v|^2)t - \\
      &\times\exp{\Bigl(2\im(uv^*)t\Bigr)} + d - c, & & -\frac{1}{2}(|u|^2 - |v|^2)\sin{2t}, \\
    D &= \Bigl(1 - d - c\cos{2t} + e\sin{2t}\Bigr)\times & \qquad
         D &= 1+(|u|^2 + |v|^2)t +  \\
      &\times\exp{\Bigl(2\im(uv^*)t\Bigr)} + d + c, & & +\frac{1}{2}(|u|^2 - |v|^2)\sin{2t}, \\
    E &= -2\Bigl(c\sin{2t} + e\cos{2t}\Bigr)\times & \qquad
         E &= 2(|u|^2 - |v|^2)\sin^2{t}, \\
      & \times\exp{\Bigl(2\im(uv^*)t\Bigr)} + 2e, & &
\end{aligned}
\end{equation}
where constants $c=c(u,v)$, $d=d(u,v)$ and $e=e(u,v)$ are defined as
\begin{equation}
\begin{split}
    c &= -\frac{(|u|^2 - |v|^2)\im(uv^*) - 2\re(uv^*)}{2\Bigl(1+\im^2(uv^*)\Bigr)}, \\
    d &= -\frac{|u|^2 + |v|^2}{2\im(uv^*)}, \\
    e &= \frac{|u|^2 - |v|^2 + 2\re(uv^*)\im(uv^*)}{2\Bigl(1+\im^2(uv^*)\Bigr)}.
\end{split}
\end{equation}
For both cases the coefficients $\lambda=\lambda(u,v,t)$ and $\delta=\delta(u,v,t)$ can be
represented as
\begin{equation}\label{e:d1d2}
\begin{split}
    \lambda &= \sqrt{2}\Bigl(\re{(\alpha)}\cos{t}+\im{(\alpha)}\sin{t}\Bigr)
               \exp{\Bigl(\im(uv^*)t\Bigr)}, \\
    \delta &= \sqrt{2}\Bigl(\im{(\alpha)}\cos{t} - \re{(\alpha)}\sin{t}\Bigr)
              \exp{\Bigl(\im(uv^*)t\Bigr)}.
\end{split}
\end{equation}
Let us find the parameter $\mu_0$, that characterizes the purity of the state $
w_{\alpha}(X,\mu ,\nu)$, or the so called purity parameter. In terms of tomograms it can be
written as \cite{jrlr-18-407}:
\begin{equation}
    \mu_0 = \frac{1}{2\pi}\int w_{\alpha}(X,\mu ,\nu ,t)
    w_{\alpha}(Y,-\mu ,-\nu ,t)\exp{\{i(X+Y)\}}\,dX\,dY\,d\mu\,d\nu.
\end{equation}
Evaluating this integral one obtains the following expression for the purity parameter:
\begin{equation}
    \mu_0 = \frac{1}{\sqrt{CD - E^{2}/4}}.
\end{equation}
Considering this expression one concludes that for parameters $u=\sqrt{\gamma}$ and
$v=i\sqrt{\gamma}$ the purity parameter equals to $1$, i.e. for such a choice of $u$ and $v$
the ground and coherent states of the oscillator remain pure for all moments of time. For
another choice of these parameters the purity $\mu_0$ is not equal to $1$.
\begin{figure}[ht]
\centering \mbox{\subfigure[$u=1$, $v=10i$]{\includegraphics{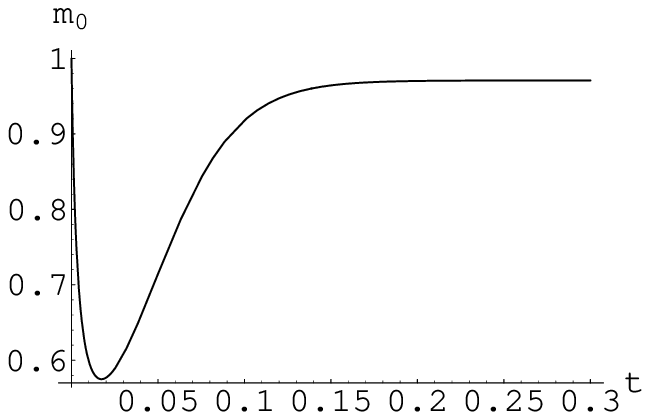}}
      \subfigure[$u=1$, $v=1+2i$]{\includegraphics{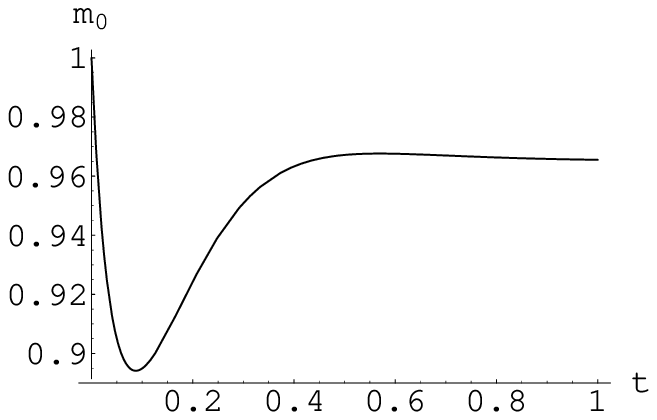}}}
\caption{Dependence of the purity on time for particular values of parameters $u$ and
$v$}\label{f:2}
\end{figure}
For example the dependence of the purity on time for some other values of parameters $u$ and
$v$ is shown on the \figurename\ \ref{f:2}. Analyzing the behavior of $\mu_{0}$ depending on
the values of $\mu$ and $\nu$ we are able to make some general conclusions. If $\im(uv^*)=0$
or $\im(uv^*)>0$ then the purity $\mu_0$ tends to the $0$ when the time $t$ tends to the
$+\infty$. If $\im(uv^*)<0$ then the purity $\mu_0$ tends to the limit
\begin{equation}
    \mu_0(u,v,+\infty) = \frac{1}{\sqrt{d^2(u,v)-c^2(u,v)-e^2(u,v)}} =
    \sqrt{\frac{4\im^2(uv^*)\Bigl(1+4\im^2(uv^*)\Bigr)}{{(|u|^2+|v|^2)}^2+16\im^4(uv^*)}}
\end{equation}
From the inequality $\im^2(uv^*) \leqslant |u|^2|v|^2$ it is obvious that this limit is always
less than unit and it is equal to the unit if $|\im(uv^*)|=|u||v|$ and $|u|=|v|$.

\section{Conclusion}

To summarize we point out the main results of this work. We reviewed the generic evolution
equation for density operator of open system in probability representation of quantum
mechanics. In this representation the state evolution is described by conventional probability
density (tomogram) evolution. In this sense the obtained equation is a generalization of Moyal
evolution equation \cite{pcps-45-99}, written for Wigner function \cite{pr-40-749}. In our
case we set the generic quantum kinetic equation for open systems similar to the Fokker-Plank
type equation. The obtained equation is written in terms of quantum state tomograms.

As example of such generic equations we have shown that the damped oscillator evolution
depends on the character of chosen values of parameters available in the collision terms of
the kinetic equation for the oscillator.

\section{Acknowledgement}

E. B. thanks the Russian Foundation for Basic Research for partial support under Proj. N
01-02-17745 and 03-02-06073.


\begin{thebibliography}{10}

\bibitem{ajp-70-288}
D. F. Styer \textit{et al}.
\newblock {\em Am. J. Phys.} \textbf{70}, 288 (2002).

\bibitem{jrlr-22-401}
V. I. Man'ko, V. A. Sharapov and E. V. Shchukin.
\newblock {\em J. Russ. Las. Res.} \textbf{22}, 401 (2002).

\bibitem{pl-a-309-176}
V. I. Man'ko, V. A. Sharapov and E. V. Shchukin.
\newblock {\em Phys. Lett. A} \textbf{309}, 176 (2003).

\bibitem{np-a-645-376}
G. G. Adamjan, N. V. Antonenko and W. Scheid.
\newblock {\em Nucl. Phys. A} \textbf{645}, 376 (1999).

\bibitem{jmp-34-3887}
A. Isar, A. Sandulescu and W. Scheid.
\newblock {\em J. Math. Phys.} \textbf{34}, 3887 (1993).

\bibitem{tp-167-7}
V. V. Dodonov and V. I. Manko.
\newblock {\em Trudy PhIAN} \textbf{167}, 7 (1986).

\bibitem{tp-191-171}
V. V. Dodonov, O. V. Manko and V. I. Manko.
\newblock {\em Trudy PhIAN} \textbf{191}, 171 (1989).

\bibitem{ap-79-489}
E.~Schr{\"{o}}dihger.
\newblock {\em Ann. Phys.} \textbf{79}, 489 (1926).

\bibitem{zp-45-430}
L.~D. Landau.
\newblock {\em Z. Physik} \textbf{45}, 430 (1927).

\bibitem{neumann-mathematische-grundlagen-quantenmechanik}
J.~von Neumann.
\newblock {\em Mathematische Grundlagen der Quantenmechanik}.
\newblock Springer, Berlin, 1932.

\bibitem{pr-121-920}
E.~C.~G. Sudarshan, P.~M. Mathews, and J.~Rau.
\newblock {\em Phys. Rev.} \textbf{121}, 3, 920 (1961).

\bibitem{cjm-24-520}
M.~Choi.
\newblock {\em Can. J. Math.} \textbf{24}, 3, 520 (1972).

\bibitem{kraus-states-effects-operations}
K.~Kraus.
\newblock {\em States, Effects and Operations: Fundamental Notions of Quantum
  Theory}.
\newblock Springer, 1983.

\bibitem{rpm-3-247}
A.~Kossakowski.
\newblock {\em Rep. Math. Phys.} \textbf{3}, 247 (1972).

\bibitem{cmp-46-43}
V.~Gorini and E.~C.~G. Sudarshan.
\newblock {\em Comm. Math. Phys.} \textbf{46}, 43 (1976).

\bibitem{cmp-48-119}
G.~Lindblad.
\newblock {\em Comm. Math. Phys.} \textbf{48}, 119 (1976).

\bibitem{jqe-3-348}
W. H. Louisell and J. H. Marburger.
\newblock {\em J. Quant. Electron.} \textbf{3}, 348 (1967).

\bibitem{pl-a-213-1}
S.~Mancini, V.~I. Man'ko, and P.~Tombesi.
\newblock {\em Phys. Lett. A} \textbf{213}, 1 (1996).

\bibitem{fp-27-801}
S.~Mancini, V.~I. Manko, and P.~Tombesi.
\newblock {\em Found. Phys.} \textbf{27}, 801 (1997).

\bibitem{jp-a-35-699}
O. V. Man'ko, V. I. Man'ko and G. Marmo.
\newblock {\em J. Phys. A} \textbf{35}, 699 (2001).

\bibitem{jp-a-34-8321}
M.~A. Man'ko, V.~I. Man'ko, and R.~V. Mendes.
\newblock {\em J. Phys. A} \textbf{34}, 8321 (2001).

\bibitem{pcps-45-99}
J.~E. Moyal.
\newblock {\em Proc. Cambridge Philos. Soc.} \textbf{45}, 99 (1949).

\bibitem{pr-40-749}
W.~Wigner.
\newblock {\em Phys. Rev.} \textbf{40}, 749 (1932).

\bibitem{jo-b-5-95}
S. De Nicola, R. Fedele, M. A. Man'ko and V. I. Man'ko.
\newblock {\em J. Opt. B} \textbf{5}, 95 (2003).

\bibitem{ap-237-226}
V. V. Dodonov and S. Mizrahi.
\newblock {\em Ann. Phys.} \textbf{237}, 226 (1995).

\bibitem{gelfand-shilov}
I. M. Gelfand and G. E. Shilov.
\newblock {\em Generalized functions vol. 1} Moscow (1958) [in Russian]

\bibitem{qso-7-615}
S.~Mancini, V.~I. Man'ko, and P.~Tombesi.
\newblock {\em Quantum Semiclass. Opt.} \textbf{7}, 615 (1995).

\bibitem{pr-177-1882}
K.~E. Cahill and R.~J. Glauber.
\newblock {\em Phys. Rev.} \textbf{177}, 1882 (1969).

\bibitem{rmp-13-149}
V.~Gorini, A.~Kossakowski, and E.~C.~G. Sudarshan.
\newblock {\em Rep. Math. Phys.} \textbf{13}, 149 (1978).

\bibitem{jrlr-18-407}
O.~V. Man'ko and V.~I. Man'ko.
\newblock {\em J. Russ. Laser Research} \textbf{18}, 407 (1997).

\bibitem{malkin-manko-dynamical-symmetries-coherent-states-quantum-systems}
I.~A. Malkin and V.~I. Man'ko.
\newblock {\em Dynamical Symmetries and Coherent States of Quantum Systems}.
\newblock Nauka, Moscow, 1979.
\newblock [in Russian].

\bibitem{pr-a-57-3291}
V.~I. Man'ko, L.~Rosa, and P.~Vitale.
\newblock {\em Phys. Rev. A} \textbf{57}, 3291 (1998).

\end{thebibliography}
\end{document}